\documentclass[]{revtex4}
\usepackage{graphicx}

\begin{document}

\title{The Folding Deuteron Optical Model Potentials}

\author{Xiaohua Li}
\email{lixiaohua@mail.nankai.edu.cn}
\author{Haixia An}
\author{Chonghai Cai}%
\email{haicai@nankai.edu.cn} \affiliation{Department of Physics,
Nankai University, Tianjin,
China 300071 Tel:86-22-23494154}%


\begin{abstract}
For 52 target nuclei with deuteron as projectile, we calculate the
reaction cross sections and elastic scattering angular
distributions, as well as the $\chi^2$ values for 11 kinds of
deuteron optical model potentials: our global deuteron optical
potentials and 10 folding optical potentials calculated with 2
phenomenological global nucleon optical potentials given by Koning
\textit{et al}(KD) and by Varner\textit{et al}(CH89), and 8
microscopic nucleon optical potentials with the generalized Skyrme
force parameters(GS1-6) and modified Skyrme force parameters(SKa,
SKb).

We find that for constructing the folding deuteron optical
potential, both SKa and SKb are the best Skyrme force parameters
of the microscopic nucleon optical potential proposed by Q. Shen
\textit{et al}.
\end{abstract}

\maketitle

\section{Introduction}
Deuteron is the simplest compound particle consisting of a proton
and a neutron. It  was
suggested\cite{NP84841958}\cite{NP211161960}\cite{NP612191965} that,
to the first approximation, the optical potential of a compound
particle should be the sum of the optical potentials of its
constituent nucleons, suitably averaged over their internal density
distributions in the compound particle.

Recently, H. An and C. Cai\cite{PRC730546052006} obtained a global
deuteron optical model potential for almost all target nuclei
ranging from $^{12}$C to $^{238}$U in the energy region below 183
MeV; we call it GL. It is a phenomenological optical potential
containing many adjustable parameters and can reproduce the existing
experimental data well. However, for those nuclei that lack
experimental data, especially for the nuclei that are far from the
beta-stability line, we can not with full confidence use it to
predict the reaction cross sections and elastic scattering angular
distributions, because it has no solid theoretical basis.

In this work, we calculate the deuteron optical model potential by
folding the proton and neutron optical potential obtained with the
phenomenological method and one of the microscopic approaches
proposed by Q. Shen \textit{et al}\cite{ZPA303691981} which has
analytical formula and is suit for a large amount of calculations in
nuclear data evaluation and analysis, and all of them do not contain
any free parameters in our calculation of deuteron optical
potentials, the reaction cross sections and the angular
distributions of elastic scattering. The $\chi^2$ represents the
degree of agreement between the calculated values of the reaction
and differential elastic cross sections and their experimental data.
We compare the $\chi^2$ values calculated with our global deuteron
optical potential(GL) and with ten kinds of folding deuteron optical
potentials, which correspond to ten kinds of nucleon optical
potentials: the phenomenological global optical potential given by
Koning \textit{et al}\cite{NPA7132312003}(KD)and by Varner
\textit{et al}\cite{PRT201571991}(CH89) and the microscopic optical
potential\cite{ZPA303691981} with the generalized Skyrme force
parameters (GS1-6)\cite{NPA2811661977}, and modified Skyrme force
parameters (SKa, SKb)\cite{NPA2583011976} for many nuclei. If we use
the folding deuteron optical potential which are obtained from
microscopic nucleon optical potential to calculate the reaction and
differential elastic cross sections for those nuclei with plenty of
experimental data, and obtain good theoretical results in agreement
with experimental values, we can with confidence extend it to those
nuclei which lack experimental data, because it has a reliable
theoretical basis. Q. Shen and co-works made a large amount of
calculations of neutron cross sections and angular distributions for
various nuclei and obtained rather good results in accordance with
experimental data\cite{NSE109-142}--\cite{CJNP10-187}. They found
that for neutron as projectile, the Skyrme force parameter GS2 is
the best one and the next is SKa. In this work we will find which is
the best Skyrme parameter for deuteron as projectile.

This paper is arranged as follows. In Sec. II, we provide the
folding model and deuteron ground state wave function used in
folding calculation. Sec. III gives results and discussion. Finally,
a summary is given in Sec. IV.

\section{The folding model and deuteron ground state wave function}
\subsection{The folding model}
 The folding model used for the deuteron optical potential $U_d$ is
\begin{eqnarray}
U_d(\vec{r})&=&<U_n+U_p+V_c>+\lambda_\pi^2
\,\frac{V_{so}+iW_{so}}{r} \,\frac{df_{so}(r)}{dr}
\,2\vec{S}\cdot\vec{l}\nonumber\\
&=&\int
d\vec{s}[U_n(\vec{r_n})+U_p(\vec{r_p})+V_c(\vec{r_p})]\phi_d^2(\vec{s})+\lambda_\pi^2
\,\frac{V_{so}+iW_{so}}{r} \,\frac{df_{so}(r)}{dr}
\,2\vec{S}\cdot\vec{l}
\end{eqnarray}
where
\begin{equation}
f_{so}(r)=[1+\exp((r-r_{so} \,A^{1/3})/a_{so})]^{-1}\;\;
\end{equation}
Here $U_n$ and $U_p$ are, respectively, the neutron and proton
optical potentials containing central real part, surface absorption
and volume absorption imaginary part, at half the bombarding energy
of the deuteron. $V_{so}$ and $W_{so}$ are, respectively, the real
and imaginary part of spin-orbital coupling potential of deuteron.
$\lambda_\pi$ is the Compton wave length of pion, usually
$\lambda_\pi^2=2.0 fm^2$. $V_c(r)$ is the coulomb potential. That is
to say, in this work, the folding model is only used for central
real, imaginary part and coulomb potentials. The spin-orbital
coupling potential can not be treated as the central potential in
folding model because the spin is 1/2 for neutron and proton and 1
for deuteron. Then we suppose that the spin-orbital coupling
potential as well as their parameters are the same as in
Ref.\cite{PRC730546052006}. And $\phi_d$ is the deuteron ground
state wave function used in folding integration. In the work of F.
G. Perey and G. R. Satchler\cite{NPA975151967}, they chose
Hulth$\acute{e}$n function as the deuteron ground state wave
function. In this work, we calculate it with numerical method, which
will be discussed in the following section. Here
$\vec{s}=\vec{r_n}-\vec{r_p}$, and
$\vec{r}=\frac{1}{2}({\vec{r_n}+\vec{r_p}}$) is the position of the
center of mass of the deuteron.

\subsection{The deuteron ground state wave function}
The deuteron consists of a proton and a neutron with a bing
energy of $\varepsilon_0=2.2259$MeV. The Hamiltonian of the
deuteron system can be written as:
\begin{equation}
H=\frac{p^2}{2m}+V(r)\\
\end{equation}
 where V(r) is the interaction between proton and neutron in
deuteron and m is the reduced mass of the deuteron system. We choose
V(r) as the Gauss form\cite{PR154-3}
\begin{equation}
V(r)=-\upsilon_0exp[-(\frac{r}{r_0})^2]
\end{equation}
In Ref.\cite{PR154-3}, the parameters of $\upsilon_0$ and $r_0$ are
taken as 72.15 MeV and 1.484 fm, respectively. The eigen equation of
the deuteron ground state can be written as
\begin{equation}
(H-\varepsilon_0)\phi_d(\vec{r})=0 \quad with \quad
\varepsilon_0=2.2259MeV
\end{equation}
where $\phi_d(\vec{r})$ is the deuteron ground state function. We
solve this equation with numerical method\cite{HEP-NP27-1005}. In
order to fit the experimental data of deuteron bound energy
$\varepsilon_0=2.2259$MeV, we change the parameter $\upsilon_0$ as
72.194 MeV, which is a little different from that in
Ref.\cite{PR154-3}. The wave function of the deuteron ground state
varying with r is plotted in Fig 1.

\section{RESULTS AND DISCUSSION}
Our theoretical calculation is carried out in the non-relativistic
frame, no consideration is given to the relativistic kinetics
corrections because they are usually very small when $E$$\leq$ 200
MeV (see Ref.\cite{PRC730546052006}). All experimental data used in
this work are taken from EXFOR(web address:
http://www.nndc.bnl.gov/). As for data errors, usually we take the
values given in EXFOR; in the case that the data errors are not
provided in EXFOR, we take them as $10\%$ of the corresponding
experimental data.

For a certain nuclide, the $\chi^{2}$ represents the deviation of
the calculated values from the experimental data, which is defined
as follows:
\begin{eqnarray}
\chi^2=\frac{\frac{W_{non}}{N_{non}}\sum\limits_{i=1}^{N_{non}}(\frac{\sigma_{non,i}^{th}-\sigma_{non,i}^{exp}}{\Delta
\sigma_{non,i}^{exp}}  )^2
+\frac{W_{el}}{N_{el}}\sum\limits_{i=1}^{N_{el}}\frac{1}{N_{i}}\sum\limits_{j=1}^{N_{i}}(\frac{\sigma_{el}^{th}(i,j)-\sigma_{el}^{exp}(i,j)}{\Delta
\sigma_{el}^{exp}(i,j)}  )^2                }{W_{non}+W_{el}}
\end{eqnarray}
where $\sigma_{el}^{th}(i,j)$ and $\sigma_{el}^{exp}(i,j)$ are the
theoretical and experimental differential cross sections at the j-th
angle with the i-th incidence energy, respectively. The subscript
$el$ means the data are for the elastic scattering angular
distribution. $\Delta \sigma_{el}^{exp}(i,j)$ is the error of
corresponding experimental data. $N_{i}$ is the number of angles for
the i-th incidence energy. $N_{el}$ is the number of incident energy
points of elastic scattering angular distribution for a given target
nucleus. $\sigma_{non,i}^{th}$ and $\sigma_{non,i}^{exp}$ are the
theoretical and experimental reaction cross sections at the i-th
incidence energy, respectively. The subscript $non$ means the data
are for the reaction (or nonelastic) cross sections. $\Delta
\sigma_{non,i}^{exp}$ is the error of corresponding experimental
data. $N_{non}$ is the number of incident energy points of reaction
cross sections for a given target nucleus. The $W_{el}$ and
$W_{non}$ are the weight of angular distribution of elastic
scattering and reaction cross sections, respectively. As for the
weights of $W_{el}$ and $W_{non}$, we believe that experimental data
of all nuclei are same reliable, and equal weights are applied with
$W_{el}$=2.0 and $W_{non}$=0.1 as in Ref.\cite{PRC730546052006}.

The target nuclei and the corresponding incident energies used in
calculations are showed in Table I. And we use $\chi_{nGL}^2$ to
express the $\chi^2$ calculated with the global deuteron optical
potential\cite{PRC730546052006} for the n-th nucleus, the
$\chi_{ni}^2$ for i=KD, CH89, GS1-6, SKa, SKb(see Table II-V) to
express the $\chi^2$ calculated with the folding deuteron optical
potential (see Eq.(1)) using the nucleon optical potential
parameters of Koning \textit{et al}\cite{NPA7132312003}, Varner
\textit{et al}\cite{PRT201571991}, microscopic optical potential
parameters\cite{NPA2811661977}\cite{NPA2583011976}, respectively.
The results are given in Table II-V. As for calculating the
microscopic optical potential of nucleon, we use the analytical
formula given by Q. Shen \textit{et al}\cite{ZPA303691981}.

In order to choose the best one for constructing folding deuteron
optical potentia among the ten kinds of nucleon optical potentials,
we define
\begin{equation}
\bar{\chi_{i}^2}=\frac{1}{N}\sum\limits_{n=1}^{N}\chi_{ni}^2 \quad
\quad \quad with \quad i=GL,KD,CH89,SKa,SKb,GS1-6
\end{equation}
\begin{equation}
\bar{\chi_{ni}^2}=\frac{11\chi_{ni}^2}{\sum\limits_{j}\chi_{nj}^2}
\quad \quad \quad with \quad i,j=GL,KD,CH89,SKa,SKb,GS1-6
\end{equation}
\begin{equation}
<\bar{\chi_{i}^2}>=\frac{1}{N}\sum\limits_{n=1}^{N}
\bar{\chi_{ni}^2}
 \quad \quad with \quad
i=GL,KD,CH89,SKa,SKb,GS1-6
\end{equation}
where N=52 is the total number of nuclei in Table I. For a given
target nucleus n, the $\bar{\chi_{ni}^2}$ defined in eq. (8) is the
relative $\chi^2$ of i-th optical potential among the 11 kinds of
optical potentials. For these 52 nuclei, $\bar{\chi_{i}^2}$ defined
in eq. (7) is the average value of these $\chi^2$ themselves;
$<\bar{\chi_{i}^2}>$ defined in eq. (9) is the average value of the
relative $\chi^2$, it can express more clearly than
$\bar{\chi_{i}^2}$ which is the best one among the 10 kinds of
nucleon optical potentials. The values of these $\chi_{ni}^2$ and
$\bar{\chi_{ni}^2}$ are given in Table II-V, and both the
$\bar{\chi_{i}^2}$ and $<\bar{\chi^2}_{i}>$ are given in Table VI.

From Table II-V, we can see that except for $^{6}Li$, $^{9}Be$,
$^{11}B$, $^{14}N$, $^{27}Al$, $^{112}Sn$, the value of
$\chi_{nGL}^2$ is always the minimum one among the 11 $\chi_{ni}^2$
values because the parameters in global deuteron optical potential
are obtained based on the experimental data of total reaction cross
sections and elastic scattering angular distributions of deuteron
itself. Generally speaking, the $\chi_{ni}^2$ for i=Ch89, KD, SKa,
SKb, are with small values and closed to each other for all the 52
nuclei, and the values of $\chi_{ni}^2$ for i=GS1-6 are much larger
than those of Ch89, KD, SKa, SKb for most of the 52 nuclei. From
Table VI we can clearly see that for both $\bar{\chi_{i}^2}$ and
$<\bar{\chi^2}_{i}>$, SKa and SKb are with nearly equal small values
which are less than those of KD and CH89, and the values of GS1-6
are much larger than those of SKa, SKb, KD and CH89. Therefore, we
can conclude that the microscopic nucleon optical potential with the
generalized Skyrme force parameters(GS1-6)\cite{NPA2811661977} are
not suitable for constructing folding deuteron optical potential.
And SKa and SKb are two best Skyrme force parameters of microscopic
nucleon optical potentials for constructing folding deuteron optical
potential.

For giving an intuitional display, as examples we plot the
experimental data and the theoretical values of the elastic
scattering angular distributions for the targets $^{24}Mg$ and
$^{120}Sn$ in Figs. 2, 3, respectively; and the reaction cross
sections for the targets $^{27}Al$, $^{90}Zr$, $^{120}Sn$ and
$^{208}Pb$ in Fig. 4. The theoretical values are calculated for the
global deuteron optical potential\cite{PRC730546052006} and two sets
of the folding deuteron optical potentials obtained from the
phenomenological global nucleon optical potential given by Varner
\textit{et al}\cite{PRT201571991} (CH89) and the microscopic optical
potential with modified Skyrme force parameters
SKa\cite{NPA2583011976}. From Figs. 2-4 we can see that the
calculated values for two sets of folding optical potential can
basically reproduce the experimental data. In Figs. 5, 6 we also
give the real part and imaginary part of three sets of deuteron
optical potentials at the incident energies 1, 10, 100, 200 MeV for
target $^{208}$Pb. Among the three sets of deuteron optical
potentials, there are two sets of folding optical potentials
constructed with the phenomenological global nucleon optical
potential CH89 and the microscopic nucleon optical potential with
the modified Skyrme force parameters SKa, as well as the global
deuteron optical potential(GL)\cite{PRC730546052006}. In all
figures, the solid lines represent the values calculated with GL,
the dashed lines correspond to CH89, and the dotted lines to SKa. To
see clearly, in Fig. 5, for the real part of deuteron optical
potentials, the curves at the top represent the true values, while
the others are multiplied by a factor of 2, 3 and 4, respectively;
in Fig. 6, for the imaginary part, the curves at the top represent
true values, while the others are multiplied by a factor of 10, $15$
and $20$, respectively.

\section{ SUMMARY }
In this work, for 52 target nuclei with deuteron as projectile, we
compare the $\chi_{ni}^2$, $\bar{\chi_{ni}^2}$, $\bar{\chi_{i}^2}$
and $<\bar{\chi^2}_{i}>$ values for 11 kinds of deuteron optical
model potentials: the global deuteron optical
potential\cite{PRC730546052006} and 10 folding optical potentials
calculated with 2 phenomenological global nucleon optical potentials
given by Koning \textit{et al}(KD) and by Varner\textit{et
al}(CH89), and 8 microscopic nucleon optical
potentials\cite{ZPA303691981} with the generalized Skyrme force
parameters(GS1-6) and the modified Skyrme force parameters(SKa,
SKb).

For giving an intuitional display, we plot the experimental data and
the theoretical values of the elastic scattering angular
distributions and/or the reaction cross sections of the targets
$^{24}Mg$, $^{27}Al$, $^{90}Zr$, $^{120}Sn$ and $^{208}Pb$ for three
kinds of deuteron optical potentials. We also compare the same three
kinds of deuteron optical potential themselves for target
$^{208}Pb$.

Through the comparisons, we find that for constructing folding
deuteron optical potential, the microscopic nucleon optical
potential\cite{ZPA303691981} with Skyrme force parameters SKa and
SKb are the best two because they can reproduce the experimental
data some better than the phenomenological global nucleon optical
potentials: KD and CH89. Considering that for neutron as projectile,
the microscopic nucleon optical potential Skyrme force parameters
GS2 and Ska are the best two, we conclude that Ska is the best
Skyrme force parameter of microscopic nucleon optical potential for
both neutron and deuteron as projectiles. The microscopic nucleon
optical potential has reliable theoretical basis, we can with
confidence use it to predict the data of cross sections and angular
distributions for those nuclei which lack experimental data and are
outside the applicable ranges of the global deuteron optical
potential\cite{PRC730546052006}.
\pagebreak 

\pagebreak \clearpage

\pagebreak \clearpage
\begin{table}
\caption{Incident deuteron energy for every nuclide used in the
calculations}
\begin{ruledtabular}
\renewcommand{\arraystretch}{0.8}
\begin{tabular}{cccccc}
Nucleus& Number of  & energy(MeV)& Nucleus & Number
of   & energy(MeV)\\
&energy points&&&energy points& \\
\hline
$^{6}$Li  & 25 &3-27.3     &$^{59}$Co  &1  &6.5        \\
$^{9}$Be  & 35 &1.1-160    &$^{58}$Ni  &13 &10.98-120  \\
$^{10}$B  & 1  &11.8       &$^{60}$Ni  &14 &8.9-97.4   \\
$^{11}$B  & 2  &5.5-11.8   &$^{65}$Cu  &3  &11-34.4    \\
$^{12}$C  & 16 &4-170      &$^{68}$Zn  &8  &9-80       \\
$^{13}$C  & 2  &13.7-17.7  &$^{70}$Ge  &1  &171        \\
$^{14}$N  & 12 &1.3-52     &$^{72}$Ge  &1  &171        \\
$^{16}$O  & 19 &0.98-171   &$^{74}$Ge  &1  &6.02       \\
$^{20}$Ne & 1  &52         &$^{86}$Sr  &1  &88         \\
$^{22}$Ne & 1  &52         &$^{89}$Y   &6  &15-85      \\
$^{24}$Mg & 17 &52-170     &$^{90}$Zr  &13 &5.5-183    \\
$^{27}$Al & 17 &5-160      &$^{93}$Nb  &5  &11.8-52    \\
$^{28}$Si & 18 &1.1-97.4   &$^{103}$Rh &5  &11.8-52    \\
$^{32}$S  & 5  &11.8-171   &$^{112}$Sn &1  &23         \\
$^{40}$Ar & 3  &11.8-56    &$^{116}$Sn &6  &23-183     \\
$^{39}$K  & 1  &12.8       &$^{118}$Sn &3  &14.5-56    \\
$^{40}$Ca & 10 &5-140      &$^{120}$Sn &12 &11-97.4    \\
$^{44}$Ca & 4  &9-56       &$^{124}$Sn &5  &14.5-97.4  \\
$^{48}$Ca & 7  &9-97.4     &$^{140}$Ce &1  &52         \\
$^{48}$Ti & 6  &9-52       &$^{181}$Ta &3  &11.8-52    \\
$^{50}$Ti & 4  &11.8-52    &$^{186}$W  &1  &12         \\
$^{50}$V  & 1  &171        &$^{197}$Au &6  &11-52      \\
$^{51}$V  & 4  &13.6-171   &$^{206}$Pb &2  &11.8-79.4  \\
$^{52}$Cr & 11 &4.39-34.4  &$^{208}$Pb &29 &9-140      \\
$^{54}$Fe & 9  &10-56      &$^{209}$Bi &3  &11-52      \\
$^{56}$Fe & 8  &5-56       &$^{232}$Th &3  &11.8-70    \\
\end{tabular}
\end{ruledtabular}
\end{table}

\clearpage \clearpage
\begin{table}
\caption{\label{tab:table2}$\chi_{ni}^2$ and $\bar{\chi_{ni}^2}$ for
every nuclide }
\begin{ruledtabular}
\renewcommand{\arraystretch}{0.8}
\begin{tabular}{cccccccccccc}
&$\chi_{nGL}^2$&$\chi_{nCH89}^2$&$\chi_{nKD}^2$&$\chi_{nSKa}^2$&$\chi_{nSKb}^2$&$\chi_{nGS1}^2$&$\chi_{nGS2}^2$&$\chi_{nGS3}^2$&$\chi_{nGS4}^2$&$\chi_{nGS5}^2$&$\chi_{nGS6}^2$\\
\raisebox{0.5cm}[0cm]{Nucleus  }
&$\bar{\chi^2}_{nGL}$&$\bar{\chi^2}_{nCH89}$&$\bar{\chi^2}_{nKD}$&$\bar{\chi^2}_{nSKa}$&$\bar{\chi^2}_{nSKb}$&$\bar{\chi^2}_{nGS1}$&$\bar{\chi^2}_{nGS2}$&$\bar{\chi^2}_{nGS3}$&$\bar{\chi^2}_{nGS4}$
&$\bar{\chi}^2_{nGS5}$&$\bar{\chi^2}_{nGS6}$\\ \hline
& 192.79 & 180.93 & 191.76 & 214.60 & 213.63 & 214.18 & 219.82 & 229.01 & 229.31 & 240.42 & 255.39 \\
\raisebox{0.5cm}[0cm]{$^{6}$Li  }
&0.89035    &0.8356  &0.88561 &0.99106 &0.98661 &0.98914 &1.01518 &1.05763 &1.05903 &1.11031 &1.17948 \\
& 347.55 & 324.29 & 401.32 & 358.08 & 394.70 & 788.80 & 782.92 & 586.44 & 396.64 & 385.59 & 296.67 \\
\raisebox{0.5cm}[0cm]{$^{9}$Be  }
&0.80266    &0.74893 &0.92683 &0.82698 &0.91155 &1.8217  &1.80813 &1.35437 &0.91603 &0.8905  &0.68516 \\
& 53.007 & 189.49 & 189.49 & 189.49 & 189.49 & 189.49 & 189.49 & 189.49 & 189.50 & 189.50 & 189.50 \\
\raisebox{0.5cm}[0cm]{$^{10}$B  }
&0.29933    &1.07004 &1.07004 &1.07006 &1.07006 &1.07005 &1.07005 &1.07006 &1.0701  &1.07009 &1.07011 \\
& 66.732 & 57.876 & 85.122 & 66.676 & 66.618 & 98.206 & 111.54 & 127.37 & 73.710 & 79.773 & 108.50 \\
\raisebox{0.5cm}[0cm]{$^{11}$B  }
&0.77914    &0.67575 &0.99386 &0.7785  &0.77781 &1.14663 &1.30235 &1.48709 &0.86062 &0.93141 &1.26683 \\
& 98.069 & 165.71 & 100.25 & 148.18 & 145.35 & 183.99 & 170.60 & 162.88 & 503.48 & 373.56 & 344.33 \\
\raisebox{0.5cm}[0cm]{$^{12}$C  }
&0.45016    &0.76064 &0.46017 &0.68017 &0.66718 &0.84456 &0.78309 &0.74765 &2.31108 &1.71471 &1.58058 \\
& 103.93 & 416.02 & 416.01 & 416.01 & 416.01 & 416.01 & 416.01 & 416.01 & 415.92 & 415.92 & 415.91 \\
\raisebox{0.5cm}[0cm]{$^{13}$C  }
&0.26811    &1.07328 &1.07325 &1.07324 &1.07326 &1.07326 &1.07327 &1.07326 &1.07302 &1.07303 &1.07301 \\
& 80.641 & 87.556 & 71.851 & 78.696 & 72.340 & 49.512 & 53.878 & 79.811 & 116.90 & 117.53 & 142.36 \\
\raisebox{0.5cm}[0cm]{$^{14}$N  }
&0.93268    &1.01266 &0.83101 &0.91018 &0.83668 &0.57265 &0.62314 &0.92308 &1.35202 &1.35937 &1.64653 \\
& 59.762 & 82.545 & 77.150 & 144.67 & 143.76 & 210.70 & 239.24 & 275.10 & 263.72 & 300.52 & 305.93 \\
\raisebox{0.5cm}[0cm]{$^{16}$O  }
&0.31258    &0.43174 &0.40352 &0.75667 &0.75193 &1.10202 &1.25134 &1.43886 &1.37939 &1.57182 &1.60012 \\
& 256.80 & 923.06 & 571.15 & 368.48 & 361.51 & 1852.2 & 2050.9 & 1882.2 & 1480.0 & 1565.4 & 1279.7 \\
\raisebox{0.5cm}[0cm]{$^{20}$Ne }
&0.22434    &0.8064  &0.49897 &0.32191 &0.31582 &1.61813 &1.79166 &1.64429 &1.29297 &1.3676  &1.11793 \\
& 174.01 & 501.73 & 460.37 & 264.27 & 263.75 & 913.28 & 1109.0 & 1130.3 & 616.47 & 717.67 & 776.51 \\
\raisebox{0.5cm}[0cm]{$^{22}$Ne }
&0.27631    &0.79671 &0.73102 &0.41964 &0.41881 &1.45021 &1.76104 &1.79475 &0.97889 &1.1396  &1.23302 \\
& 17.392 & 195.36 & 59.556 & 31.376 & 27.714 & 199.57 & 197.11 & 169.11 & 629.08 & 505.87 & 233.82 \\
\raisebox{0.5cm}[0cm]{$^{24}$Mg }
&0.08443    &0.94838 &0.28912 &0.15231 &0.13453 &0.96879 &0.95686 &0.82095 &3.05384 &2.45574 &1.13506 \\
& 88.976 & 137.92 & 79.191 & 179.09 & 191.32 & 1623.6 & 1986.4 & 546.51 & 281.54 & 302.95 & 301.30 \\
\raisebox{0.5cm}[0cm]{$^{27}$Al }
&0.17114    &0.26529 &0.15232 &0.34447 &0.368   &3.12289 &3.82086 &1.05121 &0.54154 &0.58273 &0.57954 \\
& 31.529 & 44.752 & 58.603 & 93.905 & 99.431 & 483.77 & 590.94 & 273.99 & 175.02 & 211.54 & 129.23 \\
\raisebox{0.5cm}[0cm]{$^{28}$Si }
&0.15817    &0.2245  &0.29399 &0.47108 &0.49881 &2.42691 &2.9645  &1.37451 &0.878   &1.0612  &0.64831 \\
\end{tabular}
\end{ruledtabular}
\end{table}

\pagebreak \clearpage

\begin{table}
\caption{\label{tab:table2}$\chi_{ni}^2$ and $\bar{\chi_{ni}^2}$ for
every nuclide }
\begin{ruledtabular}
\renewcommand{\arraystretch}{0.8}
\begin{tabular}{cccccccccccc}
&$\chi_{nGL}^2$&$\chi_{nCH89}^2$&$\chi_{nKD}^2$&$\chi_{nSKa}^2$&$\chi_{nSKb}^2$&$\chi_{nGS1}^2$&$\chi_{nGS2}^2$&$\chi_{nGS3}^2$&$\chi_{nGS4}^2$&$\chi_{nGS5}^2$&$\chi_{nGS6}^2$\\
\raisebox{0.5cm}[0cm]{Nucleus  }
&$\bar{\chi^2}_{nGL}$&$\bar{\chi^2}_{nCH89}$&$\bar{\chi^2}_{nKD}$&$\bar{\chi^2}_{nSKa}$&$\bar{\chi^2}_{nSKb}$&$\bar{\chi^2}_{nGS1}$&$\bar{\chi^2}_{nGS2}$&$\bar{\chi^2}_{nGS3}$&$\bar{\chi^2}_{nGS4}$
&$\bar{\chi}^2_{nGS5}$&$\bar{\chi^2}_{nGS6}$\\ \hline
& 46.506 & 143.85 & 106.04 & 196.92 & 199.92 & 782.50 & 976.09 & 483.23 & 428.83 & 524.34 & 446.46 \\
\raisebox{0.5cm}[0cm]{$^{32}$S   }
&0.11802    &0.36503 &0.2691  &0.49973 &0.50732 &1.98573 &2.47699 &1.22628 &1.08822 &1.3306  &1.13297 \\
& 10.236 & 123.82 & 190.36 & 298.92 & 321.71 & 1057.3 & 1425.6 & 1104.2 & 456.56 & 632.01 & 635.08 \\
\raisebox{0.5cm}[0cm]{$^{40}$Ar  }
&0.018      &0.21771 &0.33472 &0.52561 &0.56567 &1.85916 &2.50677 &1.94159 &0.80279 &1.11129 &1.11669 \\
& 16.777 & 746.44 & 746.44 & 746.43 & 746.44 & 746.43 & 746.43 & 746.44 & 746.39 & 746.40 & 746.40 \\
\raisebox{0.5cm}[0cm]{$^{39}$K   }
&0.02467    &1.09756 &1.09756 &1.09755 &1.09755 &1.09755 &1.09755 &1.09755 &1.09748 &1.09749 &1.09749 \\
& 17.965 & 71.776 & 49.973 & 61.990 & 62.101 & 73.365 & 83.611 & 83.214 & 109.06 & 112.00 & 89.302 \\
\raisebox{0.5cm}[0cm]{$^{40}$Ca  }
&0.24266    &0.96952 &0.67501 &0.83733 &0.83884 &0.99098 &1.12939 &1.12402 &1.47313 &1.51285 &1.20626 \\
& 23.364 & 67.939 & 108.97 & 253.94 & 290.99 & 1230.0 & 1759.3 & 569.15 & 202.39 & 300.55 & 190.06 \\
\raisebox{0.5cm}[0cm]{$^{44}$Ca  }
&0.05143    &0.14957 &0.23989 &0.55904 &0.64062 &2.70789 &3.87297 &1.25297 &0.44555 &0.66166 &0.41841 \\
& 55.450 & 134.19 & 259.94 & 411.52 & 471.67 & 1228.9 & 1683.1 & 567.94 & 264.48 & 385.25 & 281.32 \\
\raisebox{0.5cm}[0cm]{$^{48}$Ca  }
&0.10619    &0.25699 &0.49781 &0.78812 &0.90331 &2.35353 &3.2233  &1.08767 &0.50651 &0.7378  &0.53877 \\
& 5.0079 & 29.856 & 73.242 & 135.59 & 159.81 & 558.35 & 783.93 & 248.39 & 47.289 & 78.595 & 39.342 \\
\raisebox{0.5cm}[0cm]{$^{48}$Ti  }
&0.02551    &0.15209 &0.37309 &0.69071 &0.81409 &2.84422 &3.99334 &1.26529 &0.24089 &0.40037 &0.20041 \\
& 30.358 & 212.42 & 873.78 & 1574.5 & 1871.1 & 7137.7 & 9433.8 & 2633.1 & 418.30 & 943.98 & 321.91 \\
\raisebox{0.5cm}[0cm]{$^{50}$Ti  }
&0.01312    &0.09181 &0.37765 &0.6805  &0.80869 &3.08495 &4.07733 &1.13804 &0.18079 &0.40799 &0.13913 \\
& 2.0380 & 120.37 & 25.741 & 8.0949 & 8.0943 & 13.712 & 13.745 & 11.603 & 72.738 & 91.814 & 25.518 \\
\raisebox{0.5cm}[0cm]{$^{50}$V   }
&0.05698    &3.36512 &0.71964 &0.22631 &0.22629 &0.38334 &0.38425 &0.32437 &2.0335  &2.5668  &0.71341 \\
& 8.7021 & 72.869 & 272.48 & 264.24 & 305.01 & 1714.9 & 2400.2 & 503.78 & 227.03 & 361.30 & 234.37 \\
\raisebox{0.5cm}[0cm]{$^{51}$V   }
&0.01504    &0.12593 &0.47092 &0.45667 &0.52712 &2.96376 &4.14811 &0.87065 &0.39235 &0.62441 &0.40504 \\
& 18.344 & 106.05 & 106.05 & 106.05 & 106.05 & 106.05 & 106.05 & 106.05 & 106.02 & 106.03 & 106.03 \\
\raisebox{0.5cm}[0cm]{$^{52}$Cr  }
&0.18704    &1.08139 &1.08139 &1.08136 &1.08136 &1.08134 &1.08136 &1.08136 &1.08109 &1.08116 &1.08114 \\
& 7.6120 & 35.234 & 92.083 & 179.01 & 212.58 & 891.26 & 1298.1 & 358.16 & 81.396 & 135.69 & 90.474 \\
\raisebox{0.5cm}[0cm]{$^{54}$Fe  }
&0.02476    &0.11461 &0.29954 &0.58231 &0.6915  &2.8992  &4.22255 &1.16506 &0.26478 &0.44138 &0.2943  \\
& 70.286 & 246.58 & 1048.7 & 778.22 & 924.34 & 6484.5 & 9679.7 & 2476.7 & 512.00 & 894.62 & 481.51 \\
\raisebox{0.5cm}[0cm]{$^{56}$Fe  }
&0.03276    &0.11495 &0.48887 &0.36277 &0.43089 &3.0228  &4.51225 &1.15455 &0.23867 &0.41703 &0.22446 \\
\end{tabular}
\end{ruledtabular}
\end{table}

\pagebreak \clearpage

\begin{table}
\caption{\label{tab:table2}$\chi_{ni}^2$ and $\bar{\chi_{ni}^2}$ for
every nuclide }
\begin{ruledtabular}
\renewcommand{\arraystretch}{0.8}
\begin{tabular}{cccccccccccc}
&$\chi_{nGL}^2$&$\chi_{nCH89}^2$&$\chi_{nKD}^2$&$\chi_{nSKa}^2$&$\chi_{nSKb}^2$&$\chi_{nGS1}^2$&$\chi_{nGS2}^2$&$\chi_{nGS3}^2$&$\chi_{nGS4}^2$&$\chi_{nGS5}^2$&$\chi_{nGS6}^2$\\
\raisebox{0.5cm}[0cm]{Nucleus  }
&$\bar{\chi^2}_{nGL}$&$\bar{\chi^2}_{nCH89}$&$\bar{\chi^2}_{nKD}$&$\bar{\chi^2}_{nSKa}$&$\bar{\chi^2}_{nSKb}$&$\bar{\chi^2}_{nGS1}$&$\bar{\chi^2}_{nGS2}$&$\bar{\chi^2}_{nGS3}$&$\bar{\chi^2}_{nGS4}$
&$\bar{\chi}^2_{nGS5}$&$\bar{\chi^2}_{nGS6}$\\ \hline
& 12.238 & 142.16 & 142.16 & 142.16 & 142.16 & 142.16 & 142.16 & 142.16 & 142.16 & 142.16 & 142.16 \\
\raisebox{0.5cm}[0cm]{$^{59}$Co  }
&0.09388    &1.09061 &1.09061 &1.09061 &1.09061 &1.09061 &1.09061 &1.09061 &1.09061 &1.09061 &1.09061 \\
& 9.3105 & 127.99 & 688.65 & 318.95 & 348.17 & 2259.3 & 3407.3 & 781.33 & 1762.4 & 2231.7 & 555.31 \\
\raisebox{0.5cm}[0cm]{$^{58}$Ni  }
&0.0082     &0.11272 &0.60648 &0.28089 &0.30663 &1.98974 &3.0007  &0.6881  &1.55214 &1.96537 &0.48905 \\
& 14.074 & 193.92 & 277.88 & 223.37 & 258.05 & 753.79 & 1135.1 & 340.36 & 89.009 & 132.68 & 81.812 \\
\raisebox{0.5cm}[0cm]{$^{60}$Ni  }
&0.04423    &0.60944 &0.87332 &0.70201 &0.811   &2.36901 &3.56747 &1.06967 &0.27974 &0.417   &0.25712 \\
& 12.860 & 5751.0 & 5751.0 & 5750.9 & 5750.9 & 5750.9 & 5750.9 & 5750.9 & 5750.7 & 5750.8 & 5750.8 \\
\raisebox{0.5cm}[0cm]{$^{65}$Cu  }
&0.00246    &1.09977 &1.09977 &1.09976 &1.09976 &1.09976 &1.09976 &1.09976 &1.09972 &1.09974 &1.09973 \\
& 9.6331 & 56.503 & 348.34 & 344.70 & 421.49 & 2158.0 & 3523.2 & 838.89 & 127.42 & 251.82 & 85.743 \\
\raisebox{0.5cm}[0cm]{$^{68}$Zn  }
&0.01298    &0.07611 &0.46925 &0.46435 &0.56779 &2.90698 &4.74609 &1.13007 &0.17164 &0.33923 &0.1155  \\
& 6.4264 & 139.67 & 31.633 & 14.268 & 13.342 & 26.618 & 25.158 & 19.963 & 136.11 & 170.21 & 43.998 \\
\raisebox{0.5cm}[0cm]{$^{70}$Ge  }
&0.11267    &2.44874 &0.55462 &0.25016 &0.23392 &0.46669 &0.44109 &0.35001 &2.38633 &2.98436 &0.77142 \\
& 2.2557 & 419.81 & 121.76 & 17.184 & 13.203 & 75.478 & 75.501 & 44.475 & 351.02 & 439.51 & 92.181 \\
\raisebox{0.5cm}[0cm]{$^{72}$Ge  }
&0.01502    &2.79472 &0.81059 &0.1144  &0.08789 &0.50246 &0.50261 &0.29607 &2.33674 &2.92584 &0.61365 \\
& 31.177 & 335.37 & 335.37 & 335.37 & 335.37 & 335.37 & 335.37 & 335.37 & 335.37 & 335.37 & 335.37 \\
\raisebox{0.5cm}[0cm]{$^{74}$Ge  }
&0.10132    &1.08987 &1.08987 &1.08987 &1.08987 &1.08987 &1.08987 &1.08987 &1.08987 &1.08987 &1.08987 \\
& 32.495 & 572.13 & 771.57 & 53.665 & 54.013 & 218.95 & 182.90 & 94.659 & 117.53 & 110.62 & 80.547 \\
\raisebox{0.5cm}[0cm]{$^{86}$Sr  }
&0.15615    &2.74933 &3.7077  &0.25788 &0.25956 &1.05215 &0.87893 &0.45488 &0.56478 &0.53158 &0.38706 \\
& 34.956 & 210.84 & 648.50 & 449.18 & 548.08 & 2410.4 & 4229.7 & 890.80 & 158.02 & 231.94 & 167.07 \\
\raisebox{0.5cm}[0cm]{$^{89}$Y   }
&0.03853    &0.2324  &0.71481 &0.49512 &0.60412 &2.65688 &4.66225 &0.9819  &0.17417 &0.25565 &0.18416 \\
& 5.9884 & 55.942 & 210.10 & 209.68 & 228.67 & 441.86 & 620.39 & 200.14 & 113.85 & 164.23 & 120.96 \\
\raisebox{0.5cm}[0cm]{$^{90}$Zr  }
&0.02777    &0.25945 &0.9744  &0.97247 &1.06054 &2.04924 &2.87722 &0.92822 &0.52803 &0.76168 &0.56098 \\
& 6.0331 & 59.669 & 132.64 & 96.697 & 109.79 & 293.15 & 474.64 & 156.33 & 58.536 & 78.017 & 74.355 \\
\raisebox{0.5cm}[0cm]{$^{93}$Nb  }
&0.0431     &0.42625 &0.9475  &0.69076 &0.78426 &2.09412 &3.39063 &1.11676 &0.41815 &0.55732 &0.53116 \\
& 25.912 & 111.19 & 297.62 & 124.95 & 148.94 & 616.70 & 1094.4 & 249.09 & 27.319 & 41.583 & 26.426 \\
\raisebox{0.5cm}[0cm]{$^{103}$Rh }
&0.10312    &0.44248 &1.18441 &0.49725 &0.59273 &2.45422 &4.35515 &0.99128 &0.10872 &0.16548 &0.10516 \\
\end{tabular}
\end{ruledtabular}
\end{table}

\pagebreak \clearpage
\begin{table}
\caption{\label{tab:table2}$\chi_{ni}^2$ and $\bar{\chi_{ni}^2}$ for
every nuclide }
\begin{ruledtabular}
\renewcommand{\arraystretch}{0.8}
\begin{tabular}{cccccccccccc}
&$\chi_{nGL}^2$&$\chi_{nCH89}^2$&$\chi_{nKD}^2$&$\chi_{nSKa}^2$&$\chi_{nSKb}^2$&$\chi_{nGS1}^2$&$\chi_{nGS2}^2$&$\chi_{nGS3}^2$&$\chi_{nGS4}^2$&$\chi_{nGS5}^2$&$\chi_{nGS6}^2$\\
\raisebox{0.5cm}[0cm]{Nucleus  }
&$\bar{\chi^2}_{nGL}$&$\bar{\chi^2}_{nCH89}$&$\bar{\chi^2}_{nKD}$&$\bar{\chi^2}_{nSKa}$&$\bar{\chi^2}_{nSKb}$&$\bar{\chi^2}_{nGS1}$&$\bar{\chi^2}_{nGS2}$&$\bar{\chi^2}_{nGS3}$&$\bar{\chi^2}_{nGS4}$
&$\bar{\chi}^2_{nGS5}$&$\bar{\chi^2}_{nGS6}$\\ \hline
& 77.855 & 33.197 & 33.197 & 33.197 & 33.197 & 33.197 & 33.197 & 33.197 & 33.197 & 33.197 & 33.197 \\
\raisebox{0.5cm}[0cm]{$^{112}$Sn }
&2.0897     &0.89103 &0.89103 &0.89103 &0.89103 &0.89103 &0.89103 &0.89103 &0.89103 &0.89103 &0.89103 \\
& 4.6477 & 313.44 & 67.688 & 12.586 & 13.315 & 34.442 & 27.398 & 20.246 & 628.64 & 88.560 & 40.459 \\
\raisebox{0.5cm}[0cm]{$^{116}$Sn }
&0.04085    &2.7551  &0.59498 &0.11063 &0.11704 &0.30274 &0.24083 &0.17796 &5.52578 &0.77844 &0.35563 \\
& 9.4646 & 148.32 & 458.83 & 267.51 & 300.56 & 507.17 & 836.42 & 291.12 & 213.66 & 277.91 & 198.33 \\
\raisebox{0.5cm}[0cm]{$^{118}$Sn }
&0.02967    &0.46492 &1.43822 &0.83851 &0.94212 &1.58975 &2.62179 &0.91253 &0.66972 &0.8711  &0.62167 \\
& 28.745 & 144.11 & 500.76 & 206.60 & 238.92 & 761.11 & 1240.7 & 324.49 & 162.90 & 195.20 & 163.12 \\
\raisebox{0.5cm}[0cm]{$^{120}$Sn }
&0.07971    &0.39962 &1.38864 &0.57293 &0.66255 &2.11064 &3.44068 &0.89984 &0.45174 &0.5413  &0.45234 \\
& 13.999 & 41.086 & 81.178 & 86.983 & 100.78 & 130.79 & 180.37 & 75.561 & 50.654 & 76.368 & 59.802 \\
\raisebox{0.5cm}[0cm]{$^{124}$Sn }
&0.17156    &0.50352 &0.99486 &1.066   &1.23513 &1.6029  &2.21043 &0.92602 &0.62078 &0.93591 &0.73289 \\
& 54.654 & 1990.5 & 3489.8 & 812.98 & 843.26 & 1101.4 & 1355.3 & 1267.8 & 922.97 & 958.50 & 1184.6 \\
\raisebox{0.5cm}[0cm]{$^{140}$Ce }
&0.043      &1.566   &2.74556 &0.63961 &0.66343 &0.8665  &1.06627 &0.99741 &0.72615 &0.7541  &0.93197 \\
& 13.069 & 75.215 & 105.41 & 388.21 & 193.25 & 343.66 & 515.50 & 427.46 & 255.40 & 267.21 & 375.11 \\
\raisebox{0.5cm}[0cm]{$^{181}$Ta }
&0.04857    &0.27956 &0.39179 &1.44291 &0.71826 &1.27734 &1.91603 &1.58881 &0.94928 &0.9932  &1.39424 \\
& 1.0145 & 1244.7 & 1244.7 & 1244.7 & 1244.7 & 1244.7 & 1244.7 & 1244.7 & 1244.7 & 1244.7 & 1244.7 \\
\raisebox{0.5cm}[0cm]{$^{186}$W  }
&8.96515E-4 &1.09991 &1.09991 &1.09991 &1.09991 &1.09991 &1.09991 &1.09991 &1.09991 &1.09991 &1.09991 \\
& 7.6812 & 35.696 & 75.052 & 87.911 & 106.42 & 247.08 & 445.72 & 135.90 & 59.359 & 69.881 & 72.441 \\
\raisebox{0.5cm}[0cm]{$^{197}$Au }
&0.06291    &0.29234 &0.61465 &0.71996 &0.87156 &2.02355 &3.65036 &1.11295 &0.48614 &0.57231 &0.59327 \\
& 205.91 & 7309.3 & 7459.7 & 4733.7 & 4716.0 & 9804.2 & 9788.2 & 9489.0 & 8019.8 & 8002.3 & 8441.2 \\
\raisebox{0.5cm}[0cm]{$^{206}$Pb }
&0.02905    &1.0312  &1.05242 &0.66783 &0.66534 &1.38319 &1.38094 &1.33871 &1.13144 &1.12898 &1.1909  \\
& 15.849 & 80.832 & 113.44 & 144.37 & 161.28 & 297.72 & 437.82 & 235.44 & 152.40 & 163.79 & 182.23 \\
\raisebox{0.5cm}[0cm]{$^{208}$Pb }
&0.08782    &0.4479  &0.6286  &0.79994 &0.89367 &1.64971 &2.42601 &1.30456 &0.84449 &0.90757 &1.00974 \\
& 101.66 & 178.16 & 184.73 & 124.73 & 124.26 & 148.18 & 4999.0 & 139.32 & 131.99 & 135.59 & 142.30 \\
\raisebox{0.5cm}[0cm]{$^{209}$Bi }
&0.17446    &0.30575 &0.31702 &0.21405 &0.21325 &0.25429 &8.57874 &0.23908 &0.2265  &0.23268 &0.2442  \\
& 35.618 & 69.459 & 100.16 & 371.41 & 410.92 & 526.24 & 508.48 & 468.44 & 364.68 & 417.11 & 437.94 \\
\raisebox{0.5cm}[0cm]{$^{232}$Th }
&0.10559    &0.20592 &0.29694 &1.10109 &1.2182  &1.56009 &1.50743 &1.38873 &1.08112 &1.23657 &1.29831 \\
\end{tabular}
\end{ruledtabular}
\end{table}

\pagebreak \clearpage
\begin{table}
\caption{The values of $\bar{\chi_{i}^2},\;<\bar{\chi^2}_{i}>$}
\begin{ruledtabular}
\renewcommand{\arraystretch}{0.8}
\begin{tabular}{cccccccccccc}
 i=&GL&CH89&KD&SKa&SKb&GS1&GS2&GS3&GS4&GS5&GS6\\
\hline
$\bar{\chi_{i}^2}$    &51.1028 & 485.745 & 584.372  & 464.709  &  483.162 & 1152.13 &   1559.94  &  768.408  &  585.646  & 629.419 &   549.981 \\
$<\bar{\chi^2}_{i}>$ &0.19786 & 0.79667 & 0.80986  & 0.67989  &  0.70624 & 1.61375 &   2.30155  &  1.05254  &  1.02148  &  1.0355 &   0.79799 \\
\end{tabular}
\end{ruledtabular}
\end{table}

\pagebreak \clearpage

\begin{figure}[!htb]
\begin{center}
\includegraphics[width=12cm]{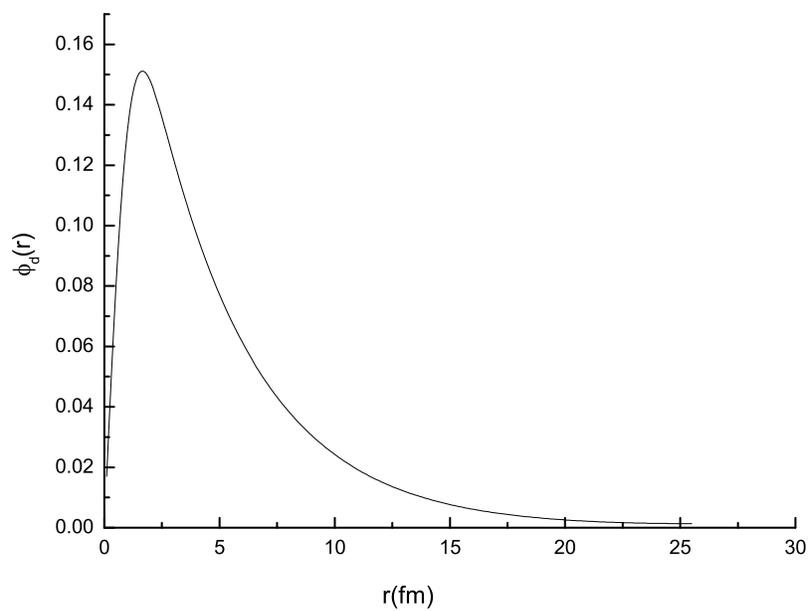}
\caption{The deuteron ground state wave function $\phi_d(r)$.}
\end{center}
\end{figure}
\begin{figure}[!htb]
\begin{center}
\includegraphics[width=12cm]{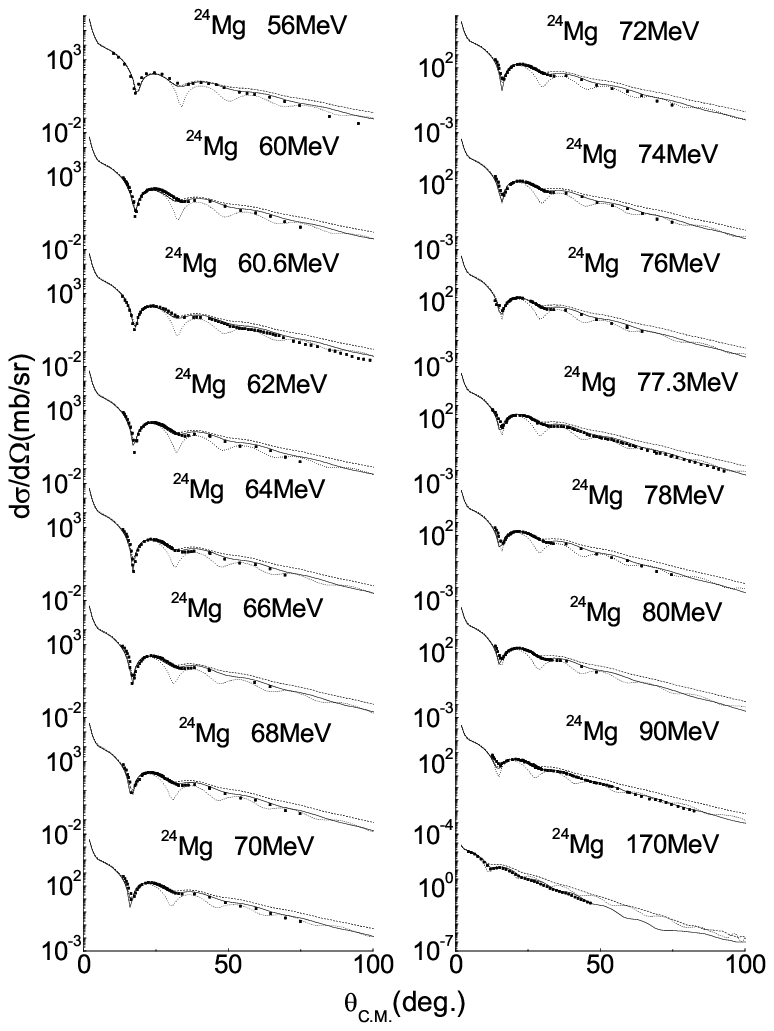}
\caption{Comparisons between the experimental angular distributions
of elastic scattering in the center of mass frame for $^{24}Mg$ and
the calculated values from the global deuteron optical potential
parameters (GL) and the folding deuteron optical potential
constructed with phenomenological global nucleon optical potential
CH89 and the microscopic nucleon optical potential with the modified
Skyrme force parameter SKa. The black dots denote the experimental
data, the solid lines represent the values of GL, the dashed lines
correspond to the values of CH89, the dotted lines to the values of
SKa. The experimental data are taken from
Refs.\cite{NPA340931980}\cite{NPA26211976}\cite{PRC630376012001},
the same symbols are used in other figures. }
\end{center}
\end{figure}

\begin{figure}[!htb]
\begin{center}
\includegraphics[width=12cm]{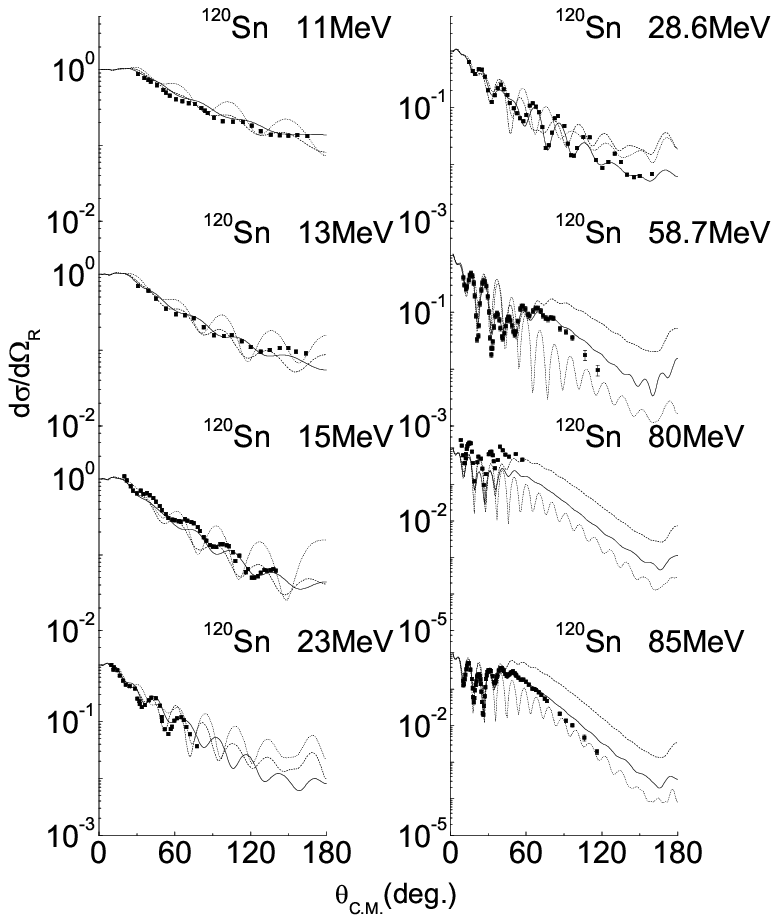}
\caption{Comparisons between the experimental angular distributions
of elastic scattering in the center of mass frame for $^{120}Sn$ and
the calculated values from the global deuteron optical potential
parameters (GL) and the folding deuteron optical potential
constructed with phenomenological global nucleon optical potential
CH89 and the microscopic nucleon optical potential with the modified
Skyrme force parameter SKa. The experimental data are taken from
Ref.\cite{NPA2323811974}\cite{PR13023911963}\cite{NPA533711991}\cite{NPA2066231973}
\cite{PRC3811531988}\cite{NPA1744851971}.}
\end{center}
\end{figure}
\begin{figure}[!htb]
\begin{center}
\includegraphics[width=12cm]{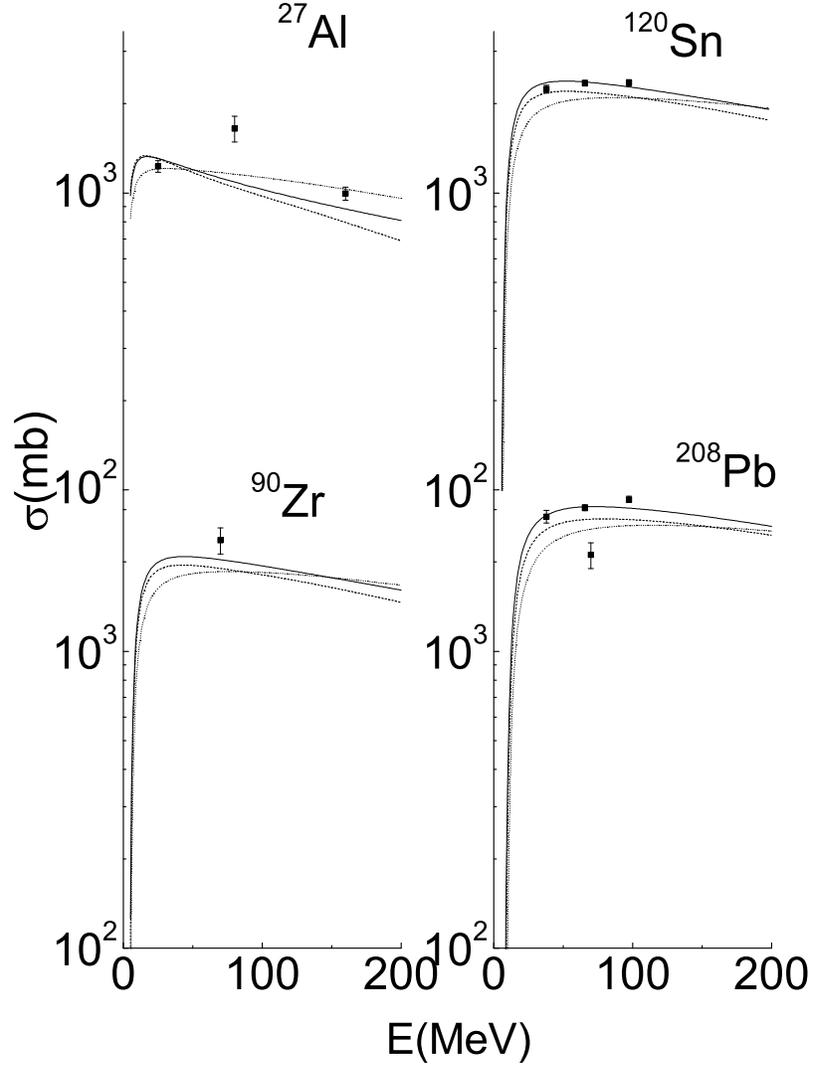}
\caption{Comparisons between the experimental reaction cross
sections and the calculated values from the global deuteron optical
potential parameters (GL) and the folding deuteron optical potential
constructed with phenomenological global nucleon optical potential
CH89 and the microscopic nucleon optical potential with the modified
Skyrme force parameter SKa. The experimental data are taken from
Ref.\cite{NP623931965}\cite{PRC193701979}\cite{PR9512681954}\cite{PRC5329191996}.}
\end{center}
\end{figure}
\begin{figure}[!htb]
\begin{center}
\includegraphics[width=12cm]{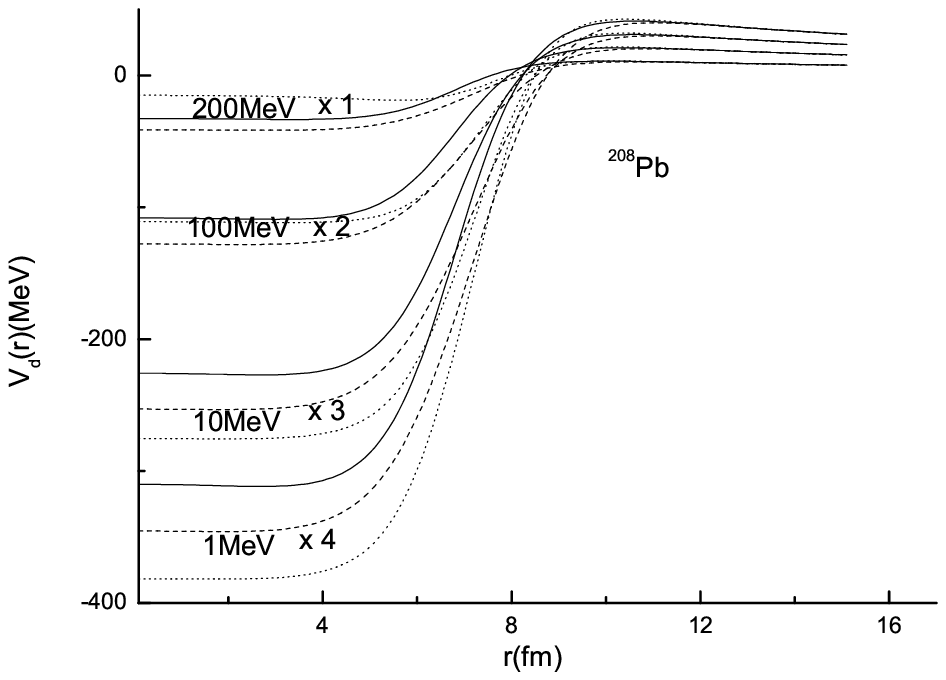}
\caption{Comparisons between the real parts of the folding deuteron
optical potentials constructed with phenomenological global nucleon
optical potential CH89 and the microscopic nucleon optical potential
with the modified Skyrme force parameter SKa and those of the global
deuteron optical potential for $^{208}Pb$.}
\end{center}
\end{figure}
\begin{figure}[!htb]
\begin{center}
\includegraphics[width=12cm]{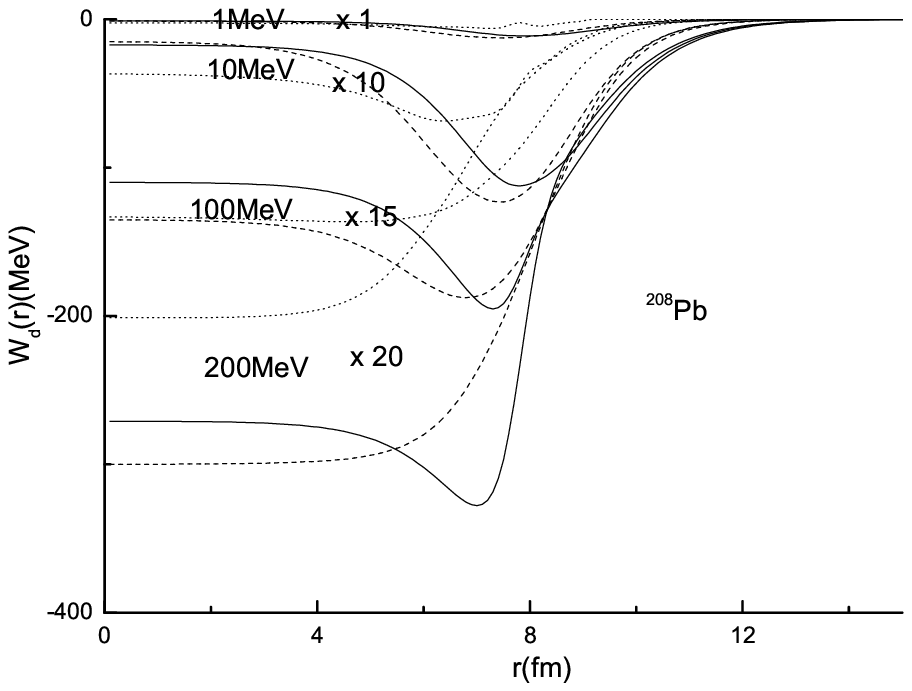}
\caption{Comparisons between the imaginary parts of the folding
deuteron optical potentials constructed with phenomenological global
nucleon optical potential CH89 and the microscopic nucleon optical
potential with the modified Skyrme force parameter SKa and those of
the global deuteron optical potential for $^{208}Pb$.}
\end{center}
\end{figure}

\pagebreak \clearpage



\clearpage







{\bf References}
\begin{thebibliography}{99}
\bibitem{NP84841958} S. Watanabe, Nucl. Phys. 8 (1958) 484.
\bibitem{NP211161960} G. R. Satchler, Nucl. Phys. 21 (1960) 116.
\bibitem{NP612191965} J. R. Rook, Nucl. Phys. 61 (1965) 219.
\bibitem{PRC730546052006} H. An, C. Cai, Phys. Rev. C, 73 (2006) 054605.
\bibitem{ZPA303691981} Q. Shen, J. Zhang, Y. Tian, Z. Ma, Y. Zhuo, Z. Phys. A 303 (1981) 69.
\bibitem{NPA7132312003} A. J. Koning, J. P. Delaroche, Nucl. Phys. A 713 (2003) 231.
\bibitem{PRT201571991} R. L. Varner, W. J. Thompson, T. L. Mcabee, E. J. Ludwig, T. B. Clegg, Phys. Rep. 201 (1991) 57.
\bibitem{NPA2811661977} S. Krewald, V. Klemt, A. Faessler, Nucl. Phys. A 281 (1977) 166.
\bibitem{NPA2583011976} H. S. Kohler, Nucl. Phys. A 258 (1976) 301.
\bibitem{NSE109-142} C. Cai, Q. Shen, Y. Zhuo, Nucl. Sci. Eng. 109 (1991) 142.
\bibitem{CJNP10-183} Y. Tian \textit{et al}, Chinese J. Nucl. Phys. 7 (1985) 154.
\bibitem{CJNP10-184} Y. Tian \textit{et al}, Chinese J. Nucl. Phys. 7 (1985) 207.
\bibitem{CJNP10-185} Y. Tian \textit{et al}, Chinese J. Nucl. Phys. 7 (1985) 344.
\bibitem{CJNP10-186} Y. Tian \textit{et al}, Chinese J. Nucl. Phys. 8 (1986) 28.
\bibitem{CJNP10-187} Y. Tian \textit{et al}, Chinese J. Nucl. Phys. 10 (1988) 183.
\bibitem{NPA975151967} F. G. Perey, G. R. Satchler, Nucl. Phys. A 97 (1967) 515.
\bibitem{PR154-3} N. Austern, Y. Iseri, M. Kamimura, M. Kawai, G. Rawitscher, M. Yahiro, Phys. Rep. 154 (1987) 125.
\bibitem{HEP-NP27-1005} C. Cai, L. Li, Chinese J. High Ener. Phys. \& Nucl. Phys.  27 (2003) 1005.


\bibitem{NPA340931980} K. Hatanaka, K. Imai, S. Kobayashi, T. Matsusue, M. Nakamura, K. Nisimura, T. Noro, H. Sakamoto, H. Shimizu, J. Shirai, Nucl. Phys. A 340 (1980) 93.
\bibitem{NPA26211976} A. Kiss, O. Aspelund, G. Hrehuss, K. T. Kn$\ddot{o}$pfle, M. Rogge, U. Schwinn, Z. Seres, P. Turek, C.
Mayer-B$\ddot{o}$ricke, Nucl. Phys. A 262 (1976) 1.
\bibitem{PRC630376012001} C. B$\ddot{a}$umer, R. Bassini, A. M. van den Berg, D. De Frenne, D. Frekers,
 M. Hagemann, V. M. Hannen, M. N. Harakeh, J. Heyse, M. A. de Huu, E. Jacobs, M. Mielke,
 S. Rakers, R. Schmidt, H. Sohlbach, H. J. W$\ddot{o}$rtche, Phys.
 Rev. C 63 (2001) 037601.
\bibitem{NPA2323811974} J. M. Lohr, W. Haeberli, Nucl. Phys. A 232 (1974) 381.
\bibitem{PR13023911963} R. K. Jolly, E. K. Lin, B. L. Cohen, Phys.
Rev. 130 (1963) 2391.
\bibitem{NPA533711991} M. Ermer, H. Clement, G. Holetzke, W. Kabitzke, G. Graw, R. Hertenberger, H. Kader, F. Merz, P. Schiemenz , Nucl. Phys. A 533 (1991) 71.
\bibitem{NPA2066231973} G. Perrin, Nguyen van Sen, J. Arvieux, C. Perrin,
R. Darves-Blanc, J. L. Durand, A. Fiore, J. C. Gondrand, F. Merchez.
Nucl. Phys. A 206 (1973) 623.
\bibitem{PRC3811531988} J. Bojowald, H. Machner, H. Nann, W. Oelert, M. Rogge, P. Turek, Phys. Rev. C 38 (1988) 1153.
\bibitem{NPA1744851971} G. Duhamel, L. Marcus, H. Langevin-Joliot, J. P. Didelez, P. Narboni, C. Stephan, Nucl. Phys. A 174 (1971) 485.
\bibitem{NP623931965} S. Mayo, W. Schimmerling, M. J. Sametband, R. M. Eisberg, Nucl. Phys. 62 (1965) 393.
\bibitem{PRC193701979} J. R. Wu, C. C. Chang, H. D. Holmgren,  Phys. Rev. C 19 (1979) 370.
\bibitem{PR9512681954} G. P. Millburn, W. Birnbaum, W. E. Crandall, L. Schecter, Phys. Rev. 95 (1954) 1268.
\bibitem{PRC5329191996} A. Auce, R. F. Carlson, A. J. Cox, A. Ingemarsson, R. Johansson, P. U. Renberg, O. Sundberg, G. Tibell, Phys. Rev. C 53 (1996) 2919.

\end{thebibliography}
\end{document}